\documentclass[a4paper,8pt,twocolumn,onecolumn]{article}%
\usepackage{amsmath}
\usepackage{amssymb}
\usepackage{amsthm}
\usepackage{graphicx}
\usepackage{subfigure}
\usepackage[latin1]{inputenc}
\usepackage{hyperref}
\usepackage{color}%
\usepackage{amsfonts}
\providecommand{\U}[1]{\protect\rule{.1in}{.1in}}
\begin{document}
\title{{\large \textbf{Dynamical Backaction Cooling with Free Electrons}}}

\author{{\normalsize A. Nigu{\`e}s$^{\mathrm{1,2}}$, A.
Siria$^{\mathrm{1,2}}$ and  P. Verlot$^{\mathrm{2,}}\footnote{Corresponding author; Email: pierre.verlot@univ-lyon1.fr}\,$}\\{\normalsize $^{\mathrm{1}}$Ecole Normale Sup{\'e}rieure,
ENS, 24, rue Lhomond 75005 Paris}\\{\normalsize $^{\mathrm{2}}$Universit{\'e} Claude Bernard Lyon 1, UCBL,}\\{\normalsize  Domaine Scientifique de La Doua, 69622 Villeurbanne, France}}

\date{\today}
\maketitle
\textbf{ The ability to cool single ions, atomic ensembles, and more recently macroscopic degrees of freedom down to the quantum groundstate has generated considerable progress and perspectives in Basic and Technological Science. These major advances have been essentially obtained by coupling mechanical motion to a resonant electromagnetic degree of freedom in what is generally known as laser cooling.  In this work, we experimentally demonstrate the first self-induced coherent cooling mechanism that is not mediated by the electromagnetic field. Using a focused electron beam, we report a 50-fold reduction of the motional temperature of a nanowire. Our result primarily relies on the sub-nanometer confinement of the electron beam and generalizes to any delayed and topologically confined interaction, with important consequences for near-field microscopy and fundamental nanoscale dissipation mechanisms.}

Coherent manipulation of mechanical motion is one of the great challenges of Modern Physics \cite{cohen1998manipulating}: Driven by such outstanding goals as reaching the zero-point motion fluctuations of single quantum objects \cite{diedrich1989laser}, observing the quantum behaviour of large atomic ensembles \cite{anderson1995observation,davis1995bose,inouye1998observation}, or engineering systems for quantum information processing \cite{kielpinski2002architecture}, a number of efficient schemes have been proposed and implemented \cite{aspect1986cooling,raab1987trapping,dalibard1989laser,kasevich1992laser}, establishing laser control as a paradigm for cooling and trapping matter at the microscopic scale \cite{metcalf1999laser}.
In recent years, the field of cavity optomechanics has demonstrated that this paradigm extends remarkably well to the macroscopic level \cite{Arcizet2006,Gigan2006}, with demonstrations of radiation-pressure induced cooling \cite{Schliesser2008b,Thomson2008} down to the quantum groundstate  \cite{chan2011laser,Teufel2011}.

Despite a considerable variability both in concepts and experimental realizations, all the above cited experiments rely on the fundamental interactions between a mechanical degree of freedom and a mode of the electromagnetic field (e.g. atomic transitions \cite{cohen1992atom}, Fabry-Perot resonance \cite{Kippenberg2008}, two-level systems \cite{Wilson-Rae2004,Arcizet2011}...) which harvests the mechanical energy. In this work, we report the first dynamical backaction cooling experiment that is not mediated by the electromagnetic field. We demonstrate that under the illumination of a continuous focused electron beam, a nanowire can spontaneously reach an equilibrium with drastically reduced  motional temperature. We develop a simple and general model and attribute this behaviour to the presence of dissipative force gradients generated by the electron-nanowire interaction.

From a general perspective, our results point out the potential of topological confinement to provide the same functions as optical confinement in laser cooling, with important consequences for interpreting and controlling near-field dynamics at the nanoscale \cite{rabe1991direct,chan2001nonlinear,rousseau2009radiative,guha2012near}. Moreover, the dramatic spatial dependence of the effectively measured mechanical damping rate emphasizes the prominent importance of taking into account the topological environment for explaining  dissipation mechanisms at the nanoscale, whose fundamental limits remain an opened question \cite{garcia2006identification,eichler2011nonlinear,cole2011phonon,rieger2014energy,nigues2014ultrahigh}. In a more specific scope, our work shows that electron microscopy is perfectly suited to ultra-sensitive, perturbation-free dynamical studies at the nanoscale, with performances comparable to laser sensing \cite{nichol2008displacement,Gloppe2014}, however with a $100$ times higher confinement. This represents a very attractive perspective for sensitive investigation of mono-dimensional structures dynamics such as carbon nanotubes \cite{baughman2002carbon} and graphene \cite{geim2007rise}.  Last, on a more technical side, our results show that electron microscopy intrinsically holds the ability to suppress the unavoidable thermal vibrations of nano-structures, yielding to a significant improvement of the image resolution.

The nano-object of interest in this work consists of a cylindrical Silicon Carbide (SiC) nanowire with length $L=150\,\mu\mathrm{m}$ and diameter $d=250\,\mathrm{nm}$. The nanowire is glued at the edge of a Tungsten micro-tip sitting on an Aluminium sample holder. The ensemble is mounted in vacuum onto the (grounded) 3D-positioning stage hosted in a commercial Scanning Electron Microscope ({{\sc Nova NanoSEM}$^\copyright$, FEI), see Figure 1(a). Scanning electron microscopy provides an image of the surface of a given sample through its response to a focused beam of electrons. The collisions between the incident electrons and the sample yield to a variety of interaction products, including light, x-rays, and electrons, which can be further detected and used for imaging purposes \cite{goldstein1981scanning}. In this study, we will focus our interest on the so-called secondary electrons (SE), which are ejected from the sample due to strongly inelastic collisions between the primary electron beam and the surface of the target. These electrons are guided to a dedicated detector (ETD, as for Everhart-Thornley Detector), which includes a strongly biased grid and a high bandwidth scintillator. Importantly, secondary emission is an absorption sensitive mechanism and therefore captures both relief and composition, making secondary electron response the most used imaging mode in Scanning Electron Microscopy (SEM). 

Here we turn secondary emission from its conventional use and show that it intrinsically holds additional capabilities for sensitive dynamical studies \cite{buks2001stiction}. The idea is depicted on Fig. 1(b): Scanning an individual nano-object in a given direction $x$, its presence is revealed under the form of a sharp peak. Its nanomechanical displacements $\delta x$ around its rest position $x_0$ will therefore result in large variations of the SE emission rate $\delta I_{\mathrm{SE}}(t)\simeq(\partial I_{\mathrm{SE}}/\partial x)_{x_0}\times \delta x(t)$, with $I_{\mathrm{SE}}(x)$ denoting the average SE emission rate as a function of position $x$. The efficiency of such a scheme is primarily determined by the secondary electron response contrast, which is typically high for a wide class of nano-objects and materials.

We have used this principle for sensitive motion detection and characterization of the SiC nanowire introduced above. All the measurements presented hereafter have been obtained with an incident electron beam current $I^{\mathrm{in}}=140\,\mathrm{pA}$ and an accelerating voltage $V=5\,\mathrm{kV}$, corresponding to an incident power $P^{\mathrm{in}}=0.7\,\mu\mathrm{W}$. Fig. 2(a) shows $2$ SEM images (with magnification coefficients of $\times 1500$ and $\times 250000$, respectively). A very high contrast can be observed, suggesting a very efficient motional transduction into the SE emission rate. To further verify this assertion, we turn the SEM into "spot mode" operation. We set the primary electron beam at position $(x_0=-100\,\mathrm{nm}, y_0=10\,\mu\mathrm{m})$ ($x$ and $y$ denote the transverse and longitudinal coordinates with respective origins taken on the axis of the nanowire and at its clamping point, see Fig. 1(a)). The SE emission rate is collected via the real-time detector output of the SEM and further sent to a spectrum analyser. Two peaks are found around a frequency $\Omega/2\pi\simeq 20\,\mathrm{kHz}$ (see Fig. 2(b)), in agreement with the theoretically expected fundamental resonance frequency $\Omega_{0,\mathrm{th}}/2\pi\simeq 0.28/L^2\sqrt{E d^2/4\rho}\simeq 18\,\mathrm{kHz}$ (SiC density $\rho=3000\,\mathrm{kg}/m^3$, and Young's modulus $E\simeq400\,\mathrm{GPa}$). We verified the presence of a pair of resonances around $\Omega/2\pi\simeq120\,\mathrm{kHz}$ (Fig. 2(c)), corresponding to approximately $6$ times the fundamental resonance frequency, and thereby completing the series of eigenmodes associated with a free-standing cantilever beam. We have also calibrated the measured spectrum into an equivalent transverse displacement (Fig. 2(b)). This is accomplished by dividing the measured fluctuations by the static transduction factor, that is the slope $(\partial I_{\mathrm{SE}}/\partial x)_{x_0,y_0}$ inferred from the line scan (inset, Fig. 2(b)). This calibration yields a displacement variance $\Delta x^2=(68\,\mathrm{pm})^2$. It is worth to compare this value to the thermal variance expected for such nanowire, $\Delta x_{\mathrm{th}}^2=k_BT/M_{\mathrm{eff}}(y_0)\Omega_{\mathrm{0}}^2$, with $k_B$ Boltzmann's constant, $T=300\,\mathrm{K}$ the ambient temperature, and $M_{\mathrm{eff}}(y_0)=0.23 M\frac{u^2(L)}{u^2(y_0)}$ the effective mass \cite{Pinard1999} ($u$ being the mode shape function associated with the fundamental flexural mode of the nanowire, and $M$ the physical mass of the nanowire, $M=\rho\times\pi d^2L/4$). For $y_0=10\,\mu\mathrm{m}$, we find $\Delta x_{\mathrm{th}}^2=(50\,\mathrm{pm})^2$, in very good agreement with the above calibrated value. We will henceforth assume the nanowire to be in contact with an external thermal bath with constant temperature $T=300\,\mathrm{K}$. Last, we have also measured spectra at various tilt angles (not shown), resulting in an effective rotation of the nanowire vibrational axis with respect to the horizontal plane. We have thus verified that the relative heights of both peaks seen on Fig.2(b) could be changed and even inverted, which we have used in order to  match both the scanning and vibrational planes.

The above reported thermal noise spectrum has been obtained within unfavourable measurement conditions, since the primary electron beam was coupled to the nanowire close to its anchor point, where  its effective mass is very large (here $M_{\mathrm{eff}}(y_0=10\,\mu\mathrm{m})=130\,\mathrm{ng}$). In a next step, we have therefore moved the electron probe towards the edge of the nanowire, expecting a rapid increase of the signal-to-noise ratio as a function of the longitudinal displacement $y$. Surprisingly, this is not what we observed: Instead, the signal-to-noise ratio remained rather constant, whereas the spectral width of the transverse mode was dramatically increased (see Fig. 2(b)).

The displacements of the nanowire are governed by the general dynamical equation:
\begin{eqnarray}
M_{\mathrm{eff}}(L)\frac{\partial^2 x}{\partial t^2}&=&-k x(t)-M_{\mathrm{eff}}(L)\Gamma_{\mathrm{M}}\frac{\partial x}{\partial t}+F_{\mathrm{th}}(t)\nonumber\\
&&+(R* F_{\mathrm{ext}}(x_{\mathrm{p}}))(t).\label{eq:1}
\end{eqnarray}
Here $x$ denotes the transverse displacement of the nanowire, $k$ its lateral spring constant, $\Gamma_{\mathrm{M}}$ its intrinsic damping rate, $F_{\mathrm{th}}(t)$ the thermal Langevin force (with spectral density $S_{\mathrm{F}}^{\mathrm{th}}[\Omega]=2M_{\mathrm{eff}}(L)\Gamma_{\mathrm{M}}k_BT$), $F_{\mathrm{ext}}$ the static external force field in which the nanowire is moving and $x_{\mathrm{p}}$ the time-dependant point of application of the force. The last term in Eq. \ref{eq:1} takes into account the external force changes resulting from the nanomechanical motion \cite{Gloppe2014}, including some possible retardation effects through a time response $R(t)$. Keeping only the time-varying component of the displacements $\delta x$ and $\delta x_{\mathrm{p}}$ and assuming that they remain small compared to the topological variations of the external force, it is straightforward to expand Eq. \ref{eq:1} in Fourier space to obtain:
\begin{eqnarray}
\chi^{-1}[\Omega]\delta x[\Omega]&=&F_{\mathrm{th}}[\Omega]+R[\Omega]\frac{\partial F_{\mathrm{ext}}}{\partial x}\Big|_{x_{\mathrm{p,eq}}}\delta x_{\mathrm{p}}[\Omega],\nonumber\\
\label{eq:2}
\end{eqnarray}
where $\Omega$ is the Fourier frequency,  $\chi[\Omega]=1/M_{\mathrm{eff}}(L)(\Omega_0^2-\Omega^2-i\Gamma_{\mathrm{M}}\Omega)$ is the mechanical susceptibility associated with the fundamental transverse vibration, and $x_{\mathrm{p,eq}}$ the static displacement of the nanowire at the point of application. In our case, the dominant external force being applied to the nanowire is generated by the primary electron beam, with point of application $(x_{\mathrm{p,eq}},y_{\mathrm{p}})$ (see notations on Fig. 2(a)). The displacement at the point of application are related to the displacement $\delta x$ through the mode shape function $u$, $\delta x_{\mathrm{p}}=\frac{u(y_{\mathrm{p}})}{u(L)}\delta x$, such that the equation of motion writes in Fourier space $\delta x[\Omega]=\chi_{\mathrm{eff}}[\Omega]F_{\mathrm{th}}[\Omega]$, with $\chi_{\mathrm{eff}}$ given by:
\begin{eqnarray}
\chi_{\mathrm{eff}}^{-1}[\Omega]&=&\chi^{-1}[\Omega]-\frac{u(y_{\mathrm{p}})}{u(L)}R[\Omega]\frac{\partial F_{\mathrm{ext}}}{\partial x}\Big|_{x_{\mathrm{p,eq}}}.\label{eq:3}
\end{eqnarray}
Assuming that the spectral variations of $R$ are negligible around frequency $\Omega_0$, Eq. \ref{eq:3} shows that our system is indeed expected to respond similarly to strongly coupled cavity optomechanical systems \cite{Arcizet2006}, whose effective mechanical response is changed in presence of cavity-delayed optical force gradients. Such delays result in an additional imaginary contribution to the effective susceptibility (the cold damping term \cite{Cohadon1999}, that would be equivalent to the imaginary part of $R$), which manifests as a change of the effective damping rate, and yields to the ability to control the dynamical state of the mechanical resonator. 

As already noted above,  we observe a very large increase of the effective dissipation when moving the probe towards its extremity, as shown on Fig. 3(b). This suggests that important delays are involved into the dynamical interaction between the primary electron beam and the nanomechanical oscillator. This led us to identify the nature of this interaction as being of a thermal origin \cite{egerton2004radiation}: When the electron beam hits the sample, a fraction of its energy is released into heat, as a consequence of inelastic mechanisms. To excite the acoustic phonon associated with the fundamental transverse vibrational mode, the produced heat needs to propagate over the entire length \cite{metzger2004cavity,rousseau2009radiative}. This propagation occurs over the heat diffusion time $\tau_h=L^2\rho c_p/\kappa$, with $c_p$ the specific heat capacity and $\kappa$ the thermal conductivity. For SiC nanowires, typical values are on the order of $c_p\simeq 750\,\mathrm{J}/(\mathrm{K}.\mathrm{kg})$ and $\kappa\simeq 10\,\mathrm{W}/(\mathrm{K}.\mathrm{m})$, the latter being a factor of $10$ lower than the bulk value, typically \cite{valentin2013comprehensive}. Hence, we have for the product $\Omega_0\tau_h\simeq640\gg1$, which places our system into a situation equivalent to the resolved sideband regime for the topological electro-thermal backaction \cite{wilson2007theory,Schliesser2008b}. This means that we can retain the dissipative contribution of the backaction force only, and write the last term of Eq. \ref{eq:3} as purely imaginary, $i(u(y_{\mathrm{p}})/u(y_{\mathrm{L}}))\times M_{\mathrm{eff}}(L)\Omega_0\Gamma_{e}(x_{\mathrm{p,eq}})$. Finally, our theoretical analysis predicts the evolution of the effective mechanical damping rate $\Gamma_{\mathrm{eff}}$ and temperature $T_{\mathrm{eff}}=\frac{M_{\mathrm{eff}}(L)\Omega_0^2(\Delta x)^2}{k_B}$  as a function of the probe longitudinal position:
\begin{eqnarray}
\Gamma_{\mathrm{eff}}(y_{\mathrm{p}})&=&\Gamma_{\mathrm{M}}+\frac{u(y_{\mathrm{p}})}{u(y_{\mathrm{L}})}\Gamma_{e}(x_{\mathrm{p,eq}}),\nonumber\\
T_{\mathrm{eff}}(y_{\mathrm{p}})&=&\frac{\Gamma_{\mathrm{M}}}{\Gamma_{\mathrm{eff}}(y_{\mathrm{p}})}\times T.\label{eq:4}
\end{eqnarray}
Note that in Eq. \ref{eq:4}, $\Gamma_{e}$, which represents the maximum dissipative coupling strength (when the probe hits the edge of the nanowire), has been assumed to only vary with the transverse degree of freedom $x_{\mathrm{p}}$ (and not $y_{\mathrm{p}}$). This is justified because of the very homogeneous secondary electron response of the nanowire (see Fig. 2(a)), which implies that the electron absorption rate (and hence the resulting thermal force) weakly depends on the longitudinal coordinate. We also emphasize the peculiar longitudinal dependence of the effective damping $u(y_{\mathrm{P}})/u(y_{\mathrm{L}})=\sqrt{M_{\mathrm{eff}}(L)/M_{\mathrm{eff}}(y_{\mathrm{P}})}$, proportional to the inverse square root of the effective mass. This is a signature of the local character of the dynamical backaction, which produces its ponderomotive effects exactly at the point of application, as demonstrated here for the first time.

Figure 3(b) shows the spectral evolution of the transverse fundamental vibration when moving the probe from the anchor to the edge of the nanowire. The whole set of data were acquired for a value of $x_{\mathrm{p,eq}}=-100\,\mathrm{nm}$. The spectra were normalized to the same effective mass, in order to better represent the dramatic reduction of the effective temperature (proportional to the spectrum area). One can also remark a slight shift of the mechanical resonance frequency at very high gains, characteristic of a residual in-phase contribution (similar to the "optical spring") in the topological backaction.  The experimental data are fitted to a Lorentzian model. The fitting parameters enable to quantitatively determine the longitudinal evolution of both the effective damping rate $\Gamma_{\mathrm{eff}}(y_{\mathrm{p}})$ and effective temperature $T_{\mathrm{eff}}(y_{\mathrm{p}})$, shown on Figs. 3(c) (black dots and green squares, respectively). Here we chose to report their evolution as functions of the effective mass at the probe location $M_{\mathrm{eff}}(y_{\mathrm{p}})$, which is a physically more representative parameter. The obtained results are in very good agreement with our theoretical description, as shown by the fitting curves (straight lines, derived from Eqs. \ref{eq:4}) which adjust very well to our experimental data. 

To further complete our study, we have also investigated the topological backaction effects when moving the equilibrium position of the probe in the transverse direction, for a fixed longitudinal coordinate $y_{\mathrm{p}}=40\,\mu\mathrm{m}$. The corresponding line scan is shown on Fig. 4 (left, dashed line). As already noted above, secondary electron response reflects the primary electrons absorption rate, which is itself proportional to the force exerted by the electron beam. As a consequence, the strength of the dynamical backaction is expected to be proportional to the gradient of the SE emission rate, $(\partial F_{\mathrm{ext}}/\partial x)\propto(\partial I_{\mathrm{SE}}/\partial x)$. Fig. 4 shows the theoretically expected backaction rate (straight, red line), obtained by normalizing the line scan derivative to the maximum backaction rate $\Gamma_{\mathrm{ba}}(y_{\mathrm{p}})=\frac{u(y_{\mathrm{p}})}{u(L)}\Gamma_{e}(x_{\mathrm{p,eq}}=-100\,\mathrm{nm})$ inferred from the measurement presented in Fig. 3(c). The right panel of Fig. 4 shows $4$ illustrative transverse positions $x_{\mathrm{p,eq}}$, labelled from $(a)$ to $(d)$. The resulting spectra show several interesting features that confirm our theoretical interpretation. Figs. 4(a) and 4(c) are both obtained on regions of positive gradient, and are showing motion sensitivities and cooling rates that are proportional to the local slope, in very good agreement with the theoretically expected backaction rates.  Fig. 4(b) is obtained with the probe being set on a gradient-free spot, resulting in the total absence of electro-mechanical transduction. Last, Fig. 4(d) corresponds to a negative slope, which conversely induces an important amplification of the Brownian motion (beyond the instability threshold, where the dynamical backaction rate cancels the intrinsic damping $\Gamma_{\mathrm{M}}$). Importantly, the reversed behaviours observed on Fig. 4(a) and 4(d) reveal the asymmetry of the backaction force with respect to the axis of the nanowire (see the cooling and heating domains on the left panel of Fig. 4). This asymmetry reflects that of the force exerted by the electron beam and is a signature of thermally-induced bending at the nanoscale \cite{ikuno2005thermally,siria2008x}.

Our study gives a quantitative access to this electro-thermal force, $F_{\mathrm{ext}}\simeq(\partial F_{\mathrm{ext}}/\partial x)\Delta x_{\mathrm{r}}\simeq M_{\mathrm{eff}}(L)\Omega_0\Gamma_e\times\Omega_{0}\tau_{\mathrm{th}}\times\Delta x_{\mathrm{r}}\simeq2\,\mathrm{pN}$, ($\Delta x_{\mathrm{r}}\simeq10\,\mathrm{nm}$ being the typical gradient range), corresponding to an equivalent static displacement of the nanowire extremity $x_{\mathrm{s}}=F_{\mathrm{ext}}/M\Omega_0^2\simeq25\,\mathrm{nm}$, in agreement with our experimental observations. This static displacement can be used in order to estimate the corresponding temperature elevation inside the nanowire \cite{ikuno2005thermally} $\Delta T\simeq dx_{\mathrm{s}}/\tilde{\alpha} L^2\simeq 70\,\mathrm{mK}$, compatible with advanced cryogenic environments (here we have taken $\tilde{\alpha}\simeq 4\times10^{-6}\,\mathrm{K}^{-1}$ the coefficient of thermal expansion of SiC). Noting $R_{\mathrm{th}}=4L/\pi d^2\kappa$ the thermal resistance of the nanowire, this temperature increase implies an energy absorption coefficient $\beta_{\mathrm{abs}}=\Delta T/R_{\mathrm{th}}P^{\mathrm{in}}=3.3\times10^{-4}$, consistent with sensitive microthermometry measurements \cite{guo2014vanadium}. 

Interestingly, the efficiency of electron beam cooling scales favourably at higher mechanical resonance frequency. Indeed, the backaction strength can be expressed as a function of the aspect ratio of the nanowire  $a=L/d$, $\Gamma_e\simeq(1/\Omega_0\tau_{\mathrm{th}})\times(1/M\Omega_0)\times M\Omega_0^2 x_{\mathrm{s}}/\Delta x_{\mathrm{r}}\propto (a/d^2)\times(\tilde{\alpha}\beta_{\mathrm{abs}}/c_p\Delta x_{\mathrm{r}})$, whereas the mechanical resonance frequency $\Omega_0\propto 1/a^2d$. Decreasing the diameter $d$ while keeping the same aspect ratio therefore yields  to both higher frequency and dynamical backaction effects. This perspective is particularly interesting in the context of coherent manipulation of ultra-low phonon number states, since the initial phonon occupancy scales as the inverse of the mechanical resonance frequency $n_0=k_BT/\hbar\Omega_0$. Hence, our scheme may provide backaction rates as high as $\Gamma_e/2\pi\simeq2\,\mathrm{MHz}$ for $\mu\mathrm{m}$-long, high aspect ratio nano-structures ($a\gtrsim 100$), that is perfectly adequate to cool them down to their quantum groundstate  \cite{moser2014nanotube}. Note that this discussion does not take into account the quantum noises associated with the electron beam, whose effects are known to limit the backaction cooling efficiency \cite{wilson2007theory,Courty2001,marquardt2007quantum}, and which remain to be quantitatively addressed both theoretically and experimentally in the present case of electron beam backaction. 

In conclusion, we have shown that free electrons establish as an ultra-sensitive, non-invasive probe for measuring and manipulating nanomechanical motion at room temperature. Using a commercial SEM, we have demonstrated that beyond its exquisite static resolution, in the nm-range, electron microscopy enables ultra-high, sub-atomic (5 pm-range) dynamical sensitivity. We have shown that the SEM appears as an active device that can be used for manipulating the dynamics of a pg-scale nanomechanical device, via an ubiquitous electro-thermal mechanism which creates strong topological gradients in the object, while weakly perturbing its static thermodynamic state. In particular, we have used this effect and reported a 50-fold suppression of the transverse vibrational mode thermal energy, representing the first self-induced cooling mechanism that is not mediated by the electromagnetic field, but entirely relies on the local force field topological confinement. Our result therefore appears as a novel, quantitative tool for ultra-sensitive study of electron matter interaction phenomena at nanoscale.
\newline

\bibliographystyle{bibliographies}
\bibliography{bib}
\paragraph{Acknowledgements} Funding for this work was provided by the PSL Excellence Chair Program.
\onecolumn
\begin{figure}[!h]
\includegraphics[width=\columnwidth]{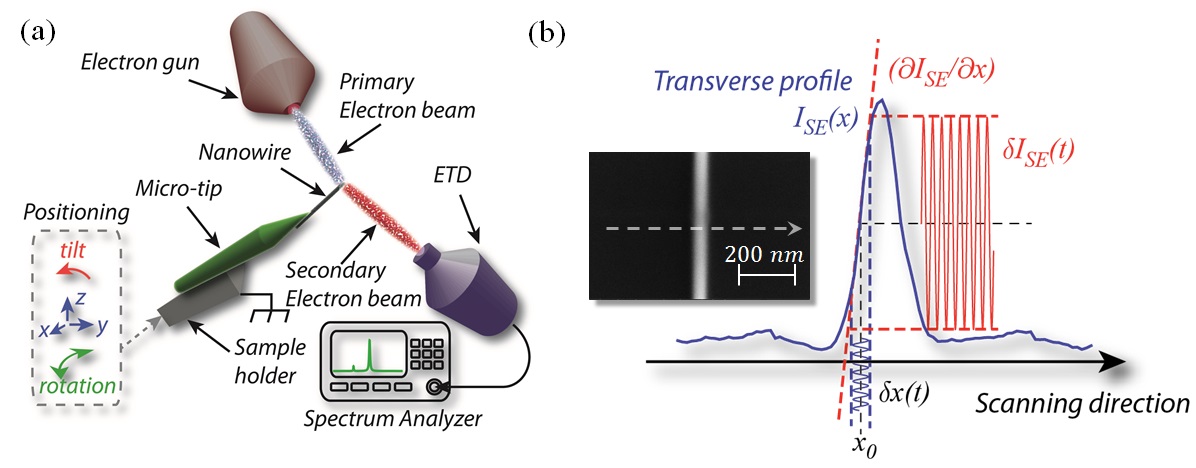} 
\centering
\caption{\textbf{Detecting Nanoscale dynamics with a SEM}. (a) Schematic of the experimental setup. The nanowire is mounted onto the 3D positioning platform of a commercial SEM, which includes an electron gun delivering a stable collimated flux of electrons, and a secondary electrons detector (ETD), whose output is used for both imaging the nanowire and measuring its dynamical motion around its equilibrium position. (b) Using Secondary Emission for nanomechanical motion detection. The very high contrast of SEM imaging (illustrated here with a $20\,\mathrm{nm}$ gold nanowire) results in a highly peaked evolution of the  SE rate as a function of the transverse displacement. The nanomechanical motion $\delta x$ around its equilibrium position $x_0$ is therefore transduced into large variation of the SE emission rate.}%
\label{Fig1}%
\end{figure}
\begin{figure}[!h]
\includegraphics[width=\columnwidth]{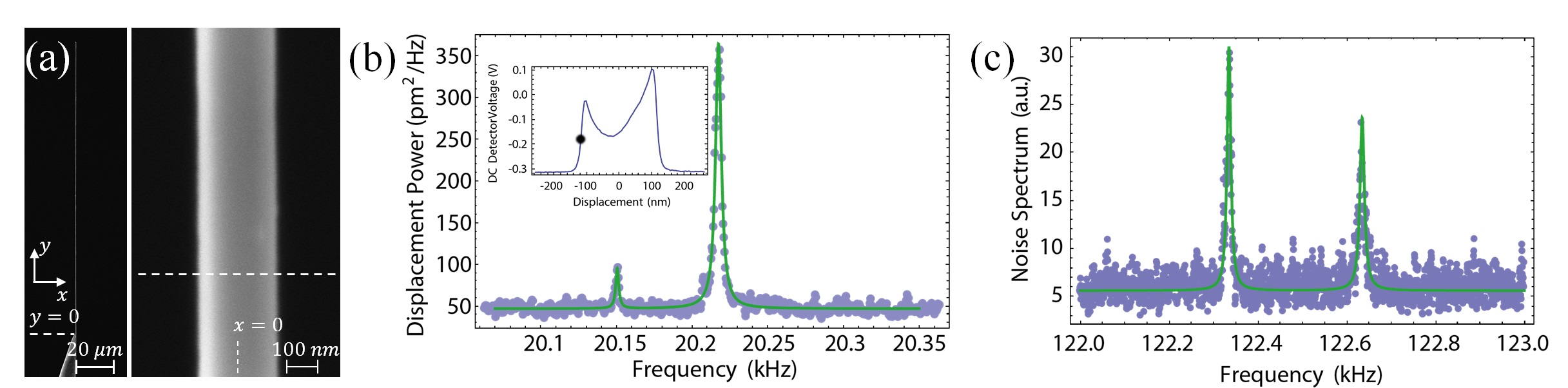} 
\centering
\caption{\textbf{Brownian motion detection of a SiC nanowire in a SEM}. (a) SEM static images of the nanowire used in the present study. The magnification coefficients are $\times 1500$ (left )and $\times 250000$ (right). The nanowire is mounted into the horizontal plane (x,y), with respective origins on the nanowire axis (right) and at the apex of the Tungsten micro-tip (left). (b) ETD noise spectrum acquired in spot mode, with the electron probe being set at $y_0=10\,\mu \mathrm{m}$. Inset shows a line scan taken at the same longitudinal distance, and which served for calibrating the spectrum into an equivalent transverse displacement. The dark dot indicates the transverse position at which the spectrum was acquired.  (c) Brownian motion spectrum associated with the second harmonic vibration around $\Omega/2\pi\simeq 122.5\,\mathrm{kHz}$.}%
\label{Fig1}%
\end{figure}
\begin{figure}[!h]
\includegraphics[width=\columnwidth]{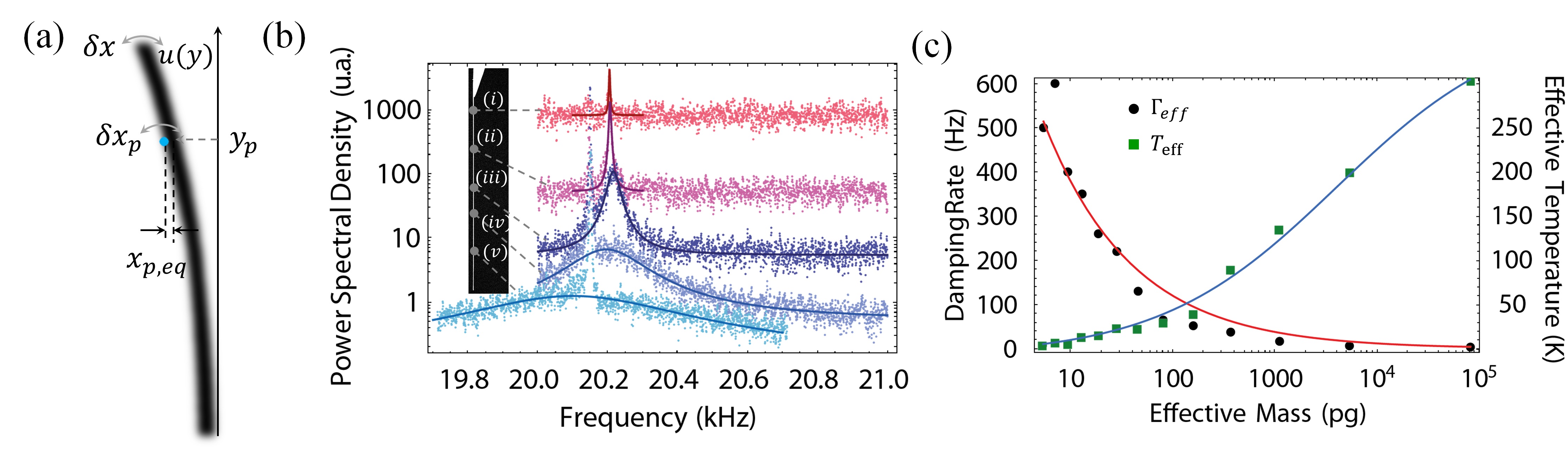} 
\centering
\caption{\textbf{Topological backaction cooling with electrons.} (a) Schematic introducing the notations used in the text. The electron probe (blue spot) is at the average position $(x_{\mathrm{p,eq}},y_{\mathrm{p}})$. The dynamical displacement  $\delta x_{\mathrm{p}}$ around the position of the probe and the tip displacement $\delta x$ are related via the mode shape function $u$ of the nanowire. (b) Topological cooling using an electron beam. The longitudinal position of the electron spot is scanned across the entire nanowire length (left), with the transverse coordinate $x_{\mathrm{p,eq}}$ being fixed. For each point, the corresponding fluctuation spectrum is recorded (acquisition time $\simeq 2\,\mathrm{min}$). The presented data are normalized to the same effective mass, in order to better visualize the drastic decrease of the effective temperature. Straight lines are a double Lorentzian adjustment of the experimental data. We attribute the observed amplification of the out-of-plane mode (leftmost peak) to the presence of orthogonal gradients. (c) Effective temperature (green squares) and effective damping rate (black dots) as functions of the effective mass. The straight lines correspond to plots of our theoretical model (Eqs. \ref{eq:4}).}%
\label{Fig2}%
\end{figure}
\begin{figure}[!h]
\includegraphics[width=\columnwidth]{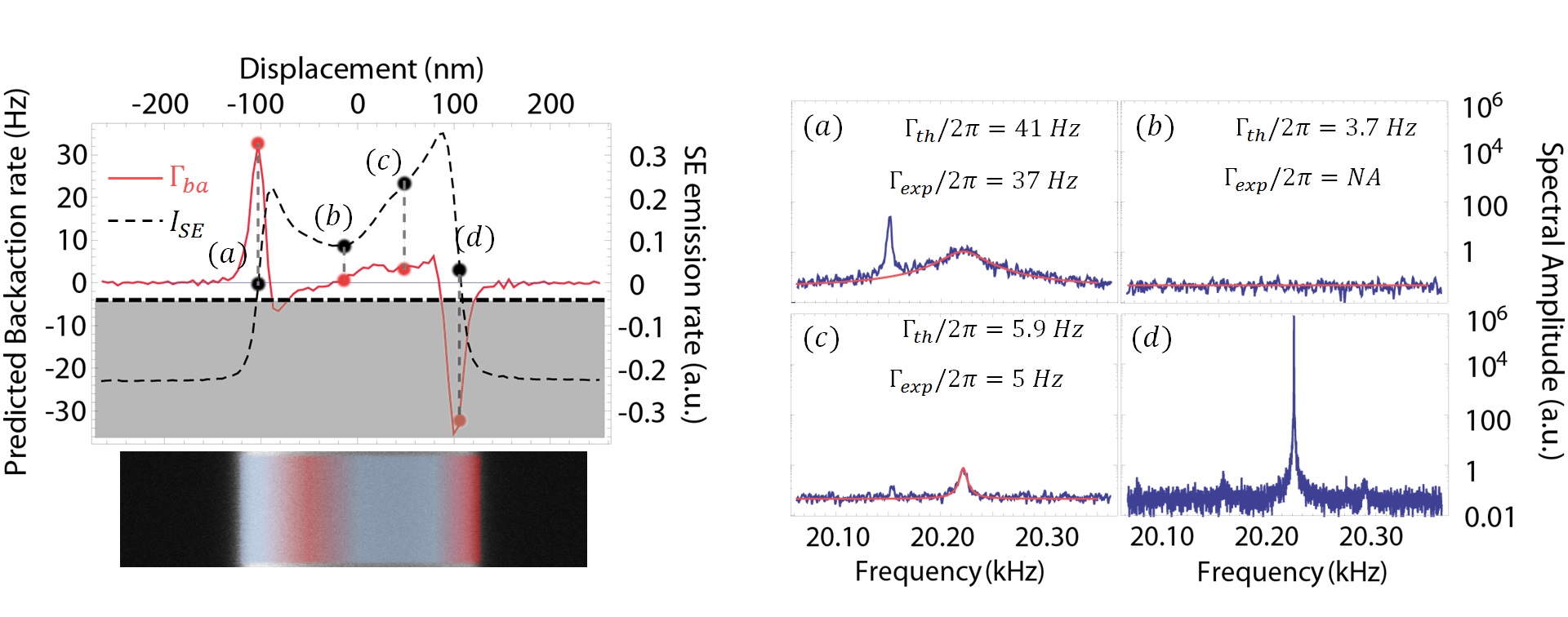} 
\centering
\caption{\textbf{Transverse evolution of the topological backaction.} Left panel: Line scans of the SE emission rate (dashed, black line) and theoretically expected backaction rate $\Gamma_{\mathrm{ba}}=\Gamma_{\mathrm{eff}}-\Gamma_{\mathrm{M}}$ (straight, red line) obtained at the longitudinal position $y_{\mathrm{p}}=40\,\mu\mathrm{m}$. The gray zone represents the parametric instability region \cite{Arcizet2006}, where the dynamical backaction cancels the intrinsic mechanical damping rate. Dots with abscissa [a-d] emphasize the values taken by $I_{\mathrm{SE}}$ and $\Gamma_{\mathrm{ba}}$ at the four acquisition spots. Heating and cooling regions are represented in red and blue on the SEM slice. Right panel: The SE emission rate fluctuation spectrum is measured at transverse positions $x_{\mathrm{p}}$ $[(a)-(d)]$. The straight lines are Lorentzian adjustments to the experimental data.}%
\label{Fig3}%
\end{figure}
\end{document}